# $K \to \pi\pi$ decays and the $K \to \sigma(600)$ weak amplitude

N. N. Trofimenkoff

E-mail: nntrof@gmail.com

February 27, 2012

**Abstract**

We use a twice-subtracted partial-wave dispersion relation in the elastic unitarity approximation for final-state interactions to study the amplitude for the $\Delta I = 1/2$ CP-conserving weak process $K + \text{spurion} \to \pi_1 + \pi_2$. We use a simple parameterization to approximate the low energy $I = 0$ S-wave $\pi\pi \to \pi\pi$ scattering phase shift and extract the residue of the $\sigma(600)$ or $f_0(600)$ pole in the weak amplitude obtained from the dispersion relation. Using this residue, we relate the $K \to \pi\pi$ decay amplitude, the weak amplitude at the soft pion point and the $K \to \sigma(600)$ weak amplitude.

## 1. Introduction

In this paper, we study the use of experimental information on isospin $I = 0$ S-wave $\pi\pi \to \pi\pi$ scattering and of a twice-subtracted partial-wave dispersion relation for a nonleptonic weak amplitude to relate the following:

(i) the $\Delta I = 1/2$ CP-conserving nonleptonic $K \to \pi_1 + \pi_2$ decay amplitude $A(M_K^2)$;

(ii) the amplitude $A(M_\pi^2)$ for this $K \to \pi_1 + \pi_2$ process at the soft pion point; and

(iii) the CP-conserving nonleptonic $K \to \sigma$ weak transition amplitude $g_{K\sigma}$, defined more precisely in Eq. (8) below, where $\sigma$ is the sigma $\sigma(600) = f_0(600)$ meson.

Such a relationship among $A(M_K^2)$, $A(M_\pi^2)$ and $g_{K\sigma}$ may be useful for providing information on any one of these three amplitudes if the other two amplitudes are obtained from experiment, theory or lattice calculations.

We use a massless spurion formalism such that $A(M_K^2)$ is converted to the amplitude for the scattering process $K + \text{spurion} \to \pi_1 + \pi_2$ with corresponding momenta $k_K + k_s \to k_1 + k_2$ and masses $M_K + 0 \to M_\pi + M_\pi$; the Mandelstam variables are $s = -(k_1 + k_2)^2$, $t = -(k_K - k_1)^2$ and $u = -(k_K - k_2)^2$ with $s + t + u = M_K^2 + 2M_\pi^2$. We denote the $S$-wave partial-wave amplitude of this scattering process by $A(s)$; then, in the limit of zero spurion momentum $k_s \to 0$, $A(s = M_K^2) = A(M_K^2)$. Also, at the soft pion point $s = M_\pi^2$, $A(s = M_\pi^2) = A(M_\pi^2)$. As specified more precisely in Eq. (8) below, $g_{K\sigma}$ is related to the residue of the sigma pole in $A(s)$ at $\sqrt{s} = \sqrt{s_\sigma} = M_\sigma - i\Gamma_\sigma/2$, where $M_\sigma$ and $\Gamma_\sigma$ are the mass and width of the sigma meson.

For our analysis, we rely extensively on some of the results of Refs. [1] and [2]. In Ref. [1], a twice-subtracted dispersion relation with a massless spurion is used to study the $K + \text{spurion} \to \pi_1 + \pi_2$ amplitude in the elastic unitarity approximation. One of their two subtraction constants is this amplitude at the soft pion point, so that a soft pion theorem may be used to relate it approximately to the nonleptonic weak transition amplitude $K \to \pi$, which may be available from lattice calculations. Their second subtraction constant is related to the slope of the amplitude at the soft pion point and is not yet readily available. The result in Ref. [1] for the $K \to \pi_1 + \pi_2$ decay amplitude depends quite strongly on this slope; therefore, information on this slope, or related equivalent information such as a knowledge of $g_{K\sigma}$, is desirable. In Ref. [2], the analysis of Ref. [1] is developed further and extended beyond the elastic unitarity approximation to include inelasticity; however, including inelasticity is beyond the scope of our paper.

This paper is organized as follows. In Sec. 2, we give our general formalism for extracting $g_{K\sigma}$ from the residue of the pole at $s = s_\sigma$ in the amplitude $A(s)$ obtained from the solution of a twice-subtracted partial-wave dispersion relation; for calculational convenience, we make one subtraction at $s = M_\pi^2$ and the second subtraction at $s = M_K^2$. Then, to simplify calculations we introduce a product form for the $I = 0$ $S$-wave $\pi\pi \to \pi\pi$ scattering $S$-matrix element to isolate the contribution from the sigma pole. In Sec. 3, as a first approximation and for mathematical simplicity, we use a simple parameterization of a portion of the $I = 0$ $S$-wave $\pi\pi \to \pi\pi$ scattering phase shift which gives the sigma pole in the $S$-matrix. Then, we evaluate the relationship among $A(M_K^2)$, $A(M_\pi^2)$ and $g_{K\sigma}$ in this first approximation. In Sec. 4, we extend our parameterization of the phase shift to reproduce, to a fairly good approximation, major features of the low energy $I = 0$ $S$-wave $\pi\pi \to \pi\pi$ scattering amplitude, and we evaluate our final result for the relationship among $A(M_K^2)$, $A(M_\pi^2)$ and $g_{K\sigma}$. In Sec. 5, we discuss our results. In Appendix A, for completeness for our formalism, we evaluate the slope of $A(s)$ at the soft pion point in terms of $A(M_\pi^2)$ and $g_{K\sigma}$, in case these become available from theoretical or lattice calculations, for possible future use in a dispersion relation with both subtractions made at $s = M_\pi^2$.

## 2. Formalism

We assume that the $S$-wave partial-wave amplitude $A(s)$ satisfies a twice-subtracted partial-wave dispersion relation with one subtraction at $s = M_\pi^2$ and the second subtraction at $s = M_K^2$ such that

$$A(s) = \frac{s - M_K^2}{M_\pi^2 - M_K^2} A(M_\pi^2) + \frac{s - M_\pi^2}{M_K^2 - M_\pi^2} A(M_K^2)$$
$$+ \frac{(s - M_\pi^2)(s - M_K^2)}{\pi} \int_{4M_\pi^2}^{\infty} \frac{\mathrm{Im} A(s') ds'}{(s' - M_\pi^2)(s' - M_K^2 - i\varepsilon)(s' - s - i\varepsilon)}$$
$$+ (L.H.C.), \tag{1}$$

where the integral represents the contribution from the right-hand unitarity cut, and *(L.H.C.)* represents the contribution from the left-hand cut. We have chosen to make one subtraction at $s = M_\pi^2$ and the second subtraction at $s = M_K^2$, rather than making both subtractions at $s = M_\pi^2$ as was done in Refs. [1] and [2], primarily for calculational convenience; our final results are independent of this different choice.

A major significant result of Ref. [1] is that the contribution from the left-hand cut is small for $s$ near $M_K^2$; we assume that it is also small for $s$ near $s_\sigma$ and neglect the term *(L.H.C)*. Also, we neglect inelasticity in $\pi\pi$ scattering, which is a good approximation below the $K\bar{K}$ threshold, and we use inelastic unitarity so that

$$\mathrm{Im} A(s) = \tan\delta(s) \mathrm{Re} A(s), \tag{2}$$

where $\delta(s)$ is the $I = 0$ $S$-wave $\pi\pi \to \pi\pi$ scattering phase shift. Then, Eq. (1) becomes

$$A(s) = \frac{s - M_K^2}{M_\pi^2 - M_K^2} A(M_\pi^2) + \frac{s - M_\pi^2}{M_K^2 - M_\pi^2} A(M_K^2)$$
$$+ \frac{(s - M_\pi^2)(s - M_K^2)}{\pi} \int_{4M_\pi^2}^{\infty} \frac{\tan\delta(s') \mathrm{Re} A(s') ds'}{(s' - M_\pi^2)(s' - M_K^2 - i\varepsilon)(s' - s - i\varepsilon)}, \tag{3}$$

which is an Omnès-Muskhelishvili singular integral equation. Eq. (3) has the solution, up to a polynomial ambiguity which we neglect,

$$A(s) = \frac{s - M_K^2}{M_\pi^2 - M_K^2} A(M_\pi^2) \Omega(s, M_\pi^2) + \frac{s - M_\pi^2}{M_K^2 - M_\pi^2} A(M_K^2) \Omega(s, M_K^2), \tag{4}$$

where $\Omega(s, s_0)$ is the once-subtracted Omnès function

$$\Omega(s, s_0) = \exp\left\{\frac{(s - s_0)}{\pi} \int_{4M_\pi^2}^{\infty} \frac{\delta(s') ds'}{(s' - s_0 - i\varepsilon)(s' - s - i\varepsilon)}\right\}, \tag{5}$$

where $i\varepsilon$ is introduced with $s_0$ for $s_0$ above the $s = 4M_\pi^2$ $\pi\pi$ scattering threshold.

To extract $g_{K\sigma}$ from the residue of the sigma pole in $A(s)$, we first define the $\sigma\pi\pi$ coupling constant $g_{\sigma\pi\pi}$ in terms of the residue $r$ of the sigma pole in the $I = 0$ S-wave $\pi\pi \to \pi\pi$ partial-wave scattering amplitude $t(s)$ by

$$g_{\sigma\pi\pi}^2 = -16\pi r \,, \qquad\qquad r = \lim_{s \to s_\sigma} (s - s_\sigma)t(s) \,, \tag{6}$$

where $t(s)$ is related to the $I = 0$ S-wave $\pi\pi \to \pi\pi$ partial-wave S-matrix element $S(s)$ by

$$S(s) = 1 + 2i\rho(s)t(s) = \exp\{2i\delta(s)\} \,, \qquad \rho(s) = \sqrt{(s - 4M_\pi^2)/s} \,. \tag{7}$$

Then, we define the CP-conserving nonleptonic $K \to \sigma$ weak transition amplitude $g_{K\sigma}$ by

$$g_{K\sigma} g_{\sigma\pi\pi} = -\lim_{s \to s_\sigma}(s - s_\sigma)A(s) \,, \tag{8}$$

since $A(s)$ has a pole in the same location as the sigma pole in $t(s)$. Combining Eqs. (4) and (8) gives

$$g_{K\sigma} = -g_{\sigma\pi\pi}^{-1} \lim_{s \to s_\sigma}(s - s_\sigma)\left\{\frac{s - M_K^2}{M_\pi^2 - M_K^2}A(M_\pi^2)\Omega(s, M_\pi^2) + \frac{s - M_\pi^2}{M_K^2 - M_\pi^2}A(M_K^2)\Omega(s, M_K^2)\right\}, \tag{9}$$

which may be rewritten as

$$g_{K\sigma} = -g_{\sigma\pi\pi}^{-1} \left(\frac{s_\sigma - M_\pi^2}{M_K^2 - M_\pi^2}\right)\left\{-\left(\frac{s_\sigma - M_K^2}{s_\sigma - M_\pi^2}\right)A(M_\pi^2)|\Omega(M_K^2, M_\pi^2)| + |A(M_K^2)|\right\}$$

$$\times \lim_{s \to s_\sigma}(s - s_\sigma)\,\Omega(s, M_K^2)\exp\{i\delta(M_K^2)\} \,. \tag{10}$$

Eq. (10) relates $A(M_K^2)$, $A(M_\pi^2)$ and $g_{K\sigma}$ for any $\delta(s)$.

To isolate the contribution from the sigma pole, we now introduce a product form for the S-matrix by separating the phase shift into two parts:

$$\delta(s) = \delta_1(s) + \delta_2(s) \,, \qquad S(s) = S_1(s)S_2(s) \,,$$

$$S_1(s) = \exp\{2i\delta_1(s)\} \,, \qquad S_2(s) = \exp\{2i\delta_2(s)\} \,,$$

$$t(s) = t_1(s)S_2(s) + t_2(s) \,, \tag{11}$$

where $t_1(s)$ or $t_2(s)$ is given by Eq. (7) with the substitution $\delta(s) \to \delta_1(s)$ or $\delta(s) \to \delta_2(s)$. Then, the Omnès function of Eq. (5) has the product form

$$\Omega(s, s_0) = \Omega_1(s, s_0)\Omega_2(s, s_0) \,, \tag{12}$$

where $\Omega_1(s, s_0)$ or $\Omega_2(s, s_0)$ is defined by Eq. (5) with the substitution $\delta(s) \to \delta_1(s)$ or $\delta(s) \to \delta_2(s)$. Henceforth, we assume that $S_1(s)$ contains the sigma pole, whereas $S_2(s)$ does not, and we define $r_1$ by

$$r_1 = \lim_{s \to s_\sigma} (s - s_\sigma) t_1(s) . \tag{13}$$

Then, Eqs. (6) and (10) can be rewritten as

$$r = r_1 S_2(s_\sigma), \qquad g_{\sigma\pi\pi}^2 = -16\pi r_1 S_2(s_\sigma), \tag{14}$$

$$g_{K\sigma} = -g_{\sigma\pi\pi}^{-1} \left(\frac{s_\sigma - M_\pi^2}{M_K^2 - M_\pi^2}\right) \left\{-\left(\frac{s_\sigma - M_K^2}{s_\sigma - M_\pi^2}\right) A(M_\pi^2) |\Omega_1(M_K^2, M_\pi^2)| |\Omega_2(M_K^2, M_\pi^2)| + |A(M_K^2)|\right\}$$

$$\times [\Omega_2(s_\sigma, M_K^2) \exp\{i\delta_2(M_K^2)\}] \times R_{\Omega 1}, \tag{15}$$

$$R_{\Omega 1} = \lim_{s \to s_\sigma} (s - s_\sigma) \Omega_1(s, M_K^2) \exp\{i\delta_1(M_K^2)\}, \tag{16}$$

because $S_2(s)$ and $\Omega_2(s, s_0)$ do not have the sigma pole. Eqs. (15) and (16) now relate $A(M_K^2)$, $A(M_\pi^2)$ and $g_{K\sigma}$ for the product form of the $S$-matrix of Eq. (11).

## 3. Parameterization of $\delta_1(s)$ and first approximation with $\delta_2(s) = 0$

For simplicity in notation and for convenience in evaluation of $\Omega(s, s_0)$, $g_{\sigma\pi\pi}$ and $g_{K\sigma}$, we introduce $\kappa = \kappa(s)$ defined by

$$\kappa = \kappa(s) = \sqrt{(s - 4M_\pi^2)/4M_\pi^2} = \rho(s)\sqrt{s/4M_\pi^2}, \tag{17}$$

which is the magnitude of the pion three-momentum divided by $M_\pi$. For our approximation, we choose the parameterization

$$\tan\delta_1(s) = \alpha\kappa(1 + \beta\kappa^2), \tag{18}$$

where $\alpha$ and $\beta$ are constants, which has been used in the study of final-state interactions in $K \to 2\pi$ decays in Refs. [3] and [4]. Eqs. (7) and (11) then give

$$S_1(s) = \frac{1 + i\alpha\kappa(1 + \beta\kappa^2)}{1 - i\alpha\kappa(1 + \beta\kappa^2)}, \qquad t_1(s) = [S_1(s) - 1]/2i\rho(s) . \tag{19}$$

Although the parameterization of Eq. (18) gives $t_1(s)$ of Eq. (19) an undesired $\sqrt{s}$ factor, it has the advantage of allowing simple evaluations of $\Omega_1(s, s_0)$ and of the residues $r_1$ and $R_{\Omega 1}$ at the sigma pole. Moreover, the constants $\alpha$ and $\beta$ can be chosen to provide an adequate first approximation of the $\pi\pi$ scattering phase shift below the $K\bar{K}$ inelastic threshold and to provide a sub-threshold or Adler zero.

If numerical values of $\alpha$ and $\beta$ are chosen first, as was done in Refs. [3] and [4], where details are provided, the poles of the $S$-matrix of Eq. (19) can be found by solving

$$1 - i\alpha\kappa(1 + \beta\kappa^2) = 0 \qquad (20)$$

by conveniently working in the complex $\kappa$ plane. For typical values of $\alpha$ and $\beta$, there are three poles in the complex $\kappa$ plane at

$$\kappa_1 = i\gamma, \qquad \kappa_2 = \mu - i\gamma/2, \qquad \kappa_3 = -\mu - i\gamma/2, \qquad (21)$$

where $\gamma$ and $\mu$ are positive real numbers, with

$$\gamma = \sqrt{4/3\beta}\ \cosh\left\{\frac{1}{3}\cosh^{-1}[(3/\alpha)/\sqrt{4/3\beta}]\right\}, \qquad \mu = \sqrt{(1/\alpha\beta\gamma) - (\gamma/2)^2}\ . \qquad (22)$$

We identify the pole $\kappa_2$ in the lower $\kappa$ plane with the sigma meson pole:

$$\kappa_\sigma = \mu - i\gamma/2, \qquad s_\sigma = 4M_\pi^2(\kappa_\sigma^2 + 1)\ . \qquad (23)$$

Conversely, if we choose numerical values of $\mu$ and $\gamma$ first, as we do in this paper, to reproduce the experimental location of the sigma pole $s_\sigma = (M_\sigma - \frac{i}{2}\Gamma_\sigma)^2$ from

$$\kappa_\sigma = \sqrt{(s_\sigma - 4M_\pi^2)/4M_\pi^2} = \mu - i\gamma/2, \qquad (24)$$

then Eqs. (18) and (19) can be used to calculate

$$\alpha = \frac{3(\gamma/2)^2 - \mu^2}{\gamma[\mu^2 + (\gamma/2)^2]}, \qquad \beta = \frac{1}{3(\gamma/2)^2 - \mu^2}\ . \qquad (25)$$

We do not ascribe any fundamental significance to the form of $\delta_1(s)$ of Eq. (18) but merely use it for mathematical simplicity in obtaining solutions which might provide some theoretical insight. Nevertheless, we note that $\delta_1(s)$ of Eq. (18) can be separated as

$$\delta_1(s) = \tan^{-1}\left\{\frac{\gamma\kappa}{[\mu^2 + (\gamma/2)^2] - \kappa^2}\right\} + \tan^{-1}\{-\kappa/\gamma\}\ , \qquad (26)$$

which shows that $\delta_1(s)$ has the form of a Breit-Wigner resonance plus a negative background phase; this is reminiscent of the "chiral shielding" recently briefly reviewed in Ref. [5]. We also note that $\tan\delta_1(s)$ of the form $\alpha\rho(s)(1 + \beta\kappa^2)$, with the factor $\rho(s)$ instead of the $\kappa$ in Eq. (18) and mentioned in relation to $K \to \pi\pi$ decays in Ref. [6], can also be separated into a (relativistic) Breit-Wigner form plus a negative background phase. However, we use Eq. (18) primarily for its much greater mathematical simplicity.

As described in Ref. [3], substituting $\delta_1(s)$ of Eq. (18) into the Omnès function of Eq. (5) yields

$$\Omega_1(s, s_0) = \frac{\kappa^2 + \gamma^2}{\kappa_0^2 + \gamma^2} \times \frac{1 - i\alpha\kappa_0(1 + \beta\kappa_0^2)}{1 - i\alpha\kappa(1 + \beta\kappa^2)}, \qquad \kappa_0 = \kappa(s = s_0) = \sqrt{(s_0 - 4M_\pi^2)/4M_\pi^2}\ . \qquad (27)$$

Since $s - s_\sigma = 4M_\pi^2(\kappa^2 - \kappa_\sigma^2)$, and since $\kappa - \kappa_\sigma$ is a factor in the denominator of $\Omega_1(s, s_0)$ of Eq. (27), the limit in Eq. (16) for $R_{\Omega 1}$ can be easily evaluated to give

$$R_{\Omega 1} = \frac{i4M_\pi^2\{\mu^2 + (\gamma/2)^2\}^2(\gamma/\mu)}{[\{(M_K^2 - 4M_\pi^2)/4M_\pi^2\} + \gamma^2]\cos\delta_1(M_K^2)} \quad . \tag{28}$$

Similarly, the limit in Eq. (13) for $r_1$ with $t_1(s)$ from Eq. (19) gives

$$r_1 = \frac{\sqrt{s_\sigma/4M_\pi^2} \times 4M_\pi^2\{\mu^2 + (\gamma/2)^2\}(\gamma/\mu)}{\mu - i\,3\gamma/2} \quad . \tag{29}$$

Eq. (15) combined with Eq. (14) and $r_1$ of Eq. (29) and with $R_{\Omega 1}$ of Eq. (28) now relates $A(M_K^2)$, $A(M_\pi^2)$ and $g_{K\sigma}$ for $\delta_1(s)$ of Eq. (18) for any $\delta_2(s)$.

As a specific numerical parameterization, we choose

$$\sqrt{s_\sigma} = M_\sigma - i\Gamma_\sigma/2 = (441^{+16}_{-8} - i272^{+9}_{-12.5}) \text{ MeV} \tag{30}$$

of Ref. [7], partly because some useful associated numerical values are readily available in papers. These associated numerical values include the $I = 0$ S-wave $\pi\pi \to \pi\pi$ scattering length $a_0^0$ from Ref. [8], the phase shift $\delta(M_K^2)_{\text{ex}}$ at $s = M_K^2$ quoted in Ref. [9], the phase shift $\delta(800)_{\text{ex}}$ at s=(800 MeV)$^2$ from Ref. [7], the location of the sub-threshold or Adler zero $s_{\text{Ad,ex}}$ from Ref. [10] and the preliminary value of the magnitude of the residue $|r|_{\text{ex}}$ of the sigma pole in $t(s)$ from Ref. [11]; we use the subscripts "ex" to avoid confusion with our calculated values. These numerical values are

$$a_0^0 = 0.220 \pm 0.005 \,, \qquad \delta(M_K^2)_{\text{ex}} = (39.2 \pm 1.5)° \,, \qquad \delta(800)_{\text{ex}} = (82.3^{+10}_{-4})° \,,$$

$$s_{\text{Ad,ex}} = (0.41 \pm 0.06)M_\pi^2 \,, \qquad |r|_{\text{ex}} = 0.218^{+0.023}_{-0.010} \text{ GeV}^2 \,; \tag{31}$$

therefore, for $\sqrt{s_\sigma}$ of Eq. (30), we can check the validity of our parameterization not only for real $s$ but also, crucially, for extrapolation of $s$ to $s_\sigma$. Then Eqs. (30), (24) and (25) give

$$\kappa_\sigma = \mu - i\gamma/2 = 1.36 - i1.14 \,, \qquad \alpha = 0.286 \,, \qquad \beta = 0.492 \,. \tag{32}$$

Then, from Eq. (32), Eqs. (18), (19) and (29) give

$$\delta_1(M_K^2) = 41.2° \,, \qquad \delta_1(800) = 74.0° \,,$$

$$s_{\text{Ad}} = -4.14 M_\pi^2 \,, \qquad r_1 = (0.207 \text{ GeV}^2)\exp(i36.6°) \,. \tag{33}$$

Since $\alpha$ would be the scattering length if $\delta_2(s)$ were zero, a comparison of Eqs. (32) and (33) with Eq. (31) shows that our numerical parameterization of $\delta_1(s)$ alone provides a promising first approximation, especially near $s = M_K^2$ and $s = s_\sigma$; as we shall see in Sec. 4, $\delta_2(s)$ can easily be chosen to reproduce $a_0^0$ and $s_{\text{Ad,ex}}$ and to improve the phase shift at s=(800 MeV)$^2$.

Eqs. (27), (28), (33) and (14) now give

$$\Omega_1(M_K^2, M_\pi^2) = (1.45)\exp(i41.2°), \qquad R_{\Omega 1} = i0.231 \text{ GeV}^2,$$

$$g_{\sigma\pi\pi}^2 = -\{(10.4 \text{ GeV}^2)\exp(i36.6°)\} S_2(s_\sigma), \tag{34}$$

so that Eq. (15) gives

$$g_{K\sigma} = -g_{\sigma\pi\pi}^{-1} \{(i0.264 \text{ GeV}^2)\exp(-i67.2°)\}$$

$$\times \{[(i1.51)\exp(i39.2°)]A(M_\pi^2)|\Omega_2(M_K^2, M_\pi^2)| + |A(M_K^2)|\}\Omega_2(s_\sigma, M_K^2)\exp\{i\delta_2(M_K^2)\}, \tag{35}$$

where we have used the numerical values of $M_\pi$, $M_K$ and $s_\sigma$. Eq. (35) with $g_{\sigma\pi\pi}$ from Eq. (34) now relates $A(M_K^2)$, $A(M_\pi^2)$ and $g_{K\sigma}$ numerically for any $\delta_2(s)$.

Before proceeding to choose $\delta_2(s)$ in Sec. 4 and because, as we shall see, our major results are not particularly sensitive to $\delta_2(s)$, we now consider the case $\delta_2(s) = 0$ for which Eqs. (34) and (35) give

$$g_{K\sigma 1} = -\{(81.8 \text{ MeV})\exp(-i85.5°)\} \times \{[(i1.51)\exp(i39.2°)]A(M_\pi^2) + |A(M_K^2)|\}, \tag{36}$$

where the subscript 1 on $g_{K\sigma 1}$ indicates that only $\delta_1(s)$ is included and $\delta_2(s) = 0$. To analyze Eq. (36) further, we note that $A(M_\pi^2)$ is expected to be much smaller than $|A(M_K^2)|$. For example, suppose we parameterize $A(M_\pi^2)$ with a parameter $y$ as

$$A(M_\pi^2) = |A(M_K^2)|/\{6(1+y)\}, \qquad -1/3 < y < 1/3; \tag{37}$$

the central value $|A(M_K^2)|/6$ is approximately typical of the results in Refs. [1] and [2]. Then, $A(M_\pi^2)$ ranges from $|A(M_K^2)|/4$, which is the result of the tree level chiral amplitude $A(s)_{\text{tree}} = A(M_\pi^2)[1 + 3(s - M_\pi^2)/(M_K^2 - M_\pi^2)]$, to $|A(M_K^2)|/8$, which encompasses the results of Refs. [1] and [2] for both elastic and inelastic final-state interactions. Then, Eq. (36) gives

$$g_{K\sigma 1}(y = 0) = -(70.6 \text{ MeV})\{\exp(-i72.4°)\}|A(M_K^2)|,$$

$$g_{K\sigma 1}(y = +1/3) = -(73.1 \text{ MeV})\{\exp(-i76.1°)\}|A(M_K^2)|,$$

$$g_{K\sigma 1}(y = -1/3) = -(66.7 \text{ MeV})\{\exp(-i64.5°)\}|A(M_K^2)|; \tag{38}$$

therefore, $g_{K\sigma 1}$ varies rather modestly around the central value $g_{K\sigma 1}(y = 0)$ as $A(M_\pi^2)$ is varied widely around the central value $|A(M_K^2)|/6$. Also, we find that $g_{K\sigma 1}$ is not very sensitive to the choice of $M_\sigma$ and $\Gamma_\sigma$; for example, if instead of Eq. (30), we use $\sqrt{s_\sigma} = \{(458 \pm 15) - i(262 \pm 15)\}$ MeV given in Ref. [12] and repeat our calculations, we get $g_{K\sigma 1}(y = 0) = -(71.9 \text{ MeV})\{\exp(-i69.0°)\}|A(M_K^2)|$ instead of the result in Eq. (38). Also, we have checked that $g_{K\sigma 1}$ is not extremely dependent on the high energy behavior of $\delta_1(s)$; as an extreme example, if we introduce a sharp cut-off at $s = s_c = (1300 \text{ MeV})^2$, where $\delta_1 = 86.1°$, by adding to $\delta_1(s)$ a phase shift $\delta_2(s) = (-90.0°)\theta(s - s_c)$, we get $g_{K\sigma 1}(y = 0) = -(74.8 \text{ MeV})\{\exp(-i69.1°)\}|A(M_K^2)|$ from Eq. (15) instead of the result in Eq. (38).

## 4. Parameterization of $\delta_2(s)$ and final results

To ameliorate some of the shortcomings of using $\delta_1(s)$ alone, we now choose the parameterization of $\delta_2(s)$ as

$$\tan\delta_2(s) = \frac{\lambda\kappa(1 + \zeta\kappa^2)}{(1 + \xi\kappa^2)} , \quad (39)$$

where $\lambda$, $\zeta$ and $\xi$ are three constants; this kind of parameterization has been used in the study of final-state interactions in $K \to \pi\pi$ decays in Ref. [4]. We choose numerical values for $\lambda$, $\zeta$ and $\xi$ as follows. The phase shift $\delta(s) = \delta_1(s) + \delta_2(s)$ resulting from $\delta_1(s)$ of Eq. (18) and $\delta_2(s)$ of Eq. (39) is given by

$$\tan\delta(s) = \frac{\alpha\kappa(1 + \beta\kappa^2)(1 + \xi\kappa^2) + \lambda\kappa(1 + \zeta\kappa^2)}{(1 + \xi\kappa^2) - \alpha\kappa(1 + \beta\kappa^2)\lambda\kappa(1 + \zeta\kappa^2)} , \quad (40)$$

which immediately gives the scattering length

$$a_0^0 = \alpha + \lambda , \quad (41)$$

so that the numerical value $a_o^o = 0.220$ of Eq. (31) can now be simply reproduced by suitably choosing $\lambda$. Next, the numerical values of the other two constants $\zeta$ and $\xi$ can be conveniently chosen by simultaneously fitting the location of the sub-threshold or Adler zero $s_{\text{Ad,exp}} = 0.41 M_\pi^2$ of Eq. (31), for which the numerator in Eq. (40) is zero, and choosing the value of $s = s_{90°}$ at which $\delta(s_{90°}) = 90°$, for which the denominator in Eq. (40) is zero. As a compromise to matching $\delta(800)_{\text{ex}}$ and $|r|_{\text{ex}}$ of Eq. (31), we choose $s_{90°} = (1100 \text{ MeV})^2$; we discuss this compromise in detail after we evaluate $r$. Using these choices, we get

$$\lambda = -0.0663 , \quad \zeta = -0.329 , \quad \xi = 0.516 ; \quad (42)$$

these numerical values yield a $\delta_2(s)$ that provides small corrections to $\delta_1(s)$ ranging from at most about $-2°$ below $s = M_K^2$ and rising to about $3°$ at $s = (800 \text{ MeV})^2$ and about $6°$ at $s = (1100 \text{ MeV})^2$.

The phase shift of Eq. (39) gives $S_2(s)$ of Eq. (11) as

$$S_2(s) = \frac{(1 + \xi\kappa^2) + i\lambda\kappa(1 + \zeta\kappa^2)}{(1 + \xi\kappa^2) - i\lambda\kappa(1 + \zeta\kappa^2)} \quad (43)$$

and the Omnès function $\Omega_2(s, s_0)$ of Eq. (12) as

$$\Omega_2(s, s_0) = \frac{\kappa^2 + \omega^2}{\kappa_0^2 + \omega^2} \times \frac{(1 + \xi\kappa_0^2) - i\lambda\kappa_0(1 + \zeta\kappa_0^2)}{(1 + \xi\kappa^2) - i\lambda\kappa(1 + \zeta\kappa^2)} \tag{44}$$

for use in Eqs. (34) and (35) to evaluate $g_{K\sigma}$. In Eq. (44), for the numbers in Eq. (42), $i\omega$ is the purely imaginary solution of

$$(1 + \xi\kappa^2) - i\lambda\kappa(1 + \zeta\kappa^2) = 0 \tag{45}$$

in the upper complex $\kappa$ plane, and the two other solutions are in the lower complex $\kappa$ plane. For the numbers in Eq. (42), the solution is

$$i\omega = i1.30, \tag{46}$$

and the other two solutions are at $-i1.51$ and $-i23.4$. Evaluating $S_2(s_\sigma)$ from Eq. (43) and $r$ and $g_{\sigma\pi\pi}$ from Eq. (14) with $r_1$ from Eq. (33) gives

$$S_2(s_\sigma) = (1.14)\exp(-i4.03°),$$

$$r = (0.236 \text{ GeV}^2)\exp(i32.6°), \qquad g_{\sigma\pi\pi}^2 = -(11.9 \text{ GeV}^2)\exp(i32.6°). \tag{47}$$

Evaluating $\Omega_2(s, s_0)$ from Eq. (44) gives

$$\Omega_2(M_K^2, M_\pi^2) = (1.06)\exp(-i0.749°), \qquad \Omega_2(s_\sigma, M_K^2) = (1.10)\exp(-i4.44°),$$

$$\Omega_2(s_\sigma, M_K^2)\exp\{i\delta_2(M_K^2)\} = (1.10)\exp(-i5.19°). \tag{48}$$

Eqs. (47) and (48) substituted into Eqs. (34) and (35) give

$$g_{K\sigma} = -\{(84.0 \text{ MeV})\exp(-i88.7°)\} \times \{[(i1.60)\exp(i39.2°)]A(M_\pi^2) + |A(M_K^2)|\}, \tag{49}$$

which is our final result for the relationship among $A(M_K^2)$, $A(M_\pi^2)$ and $g_{K\sigma}$. To facilitate comparison with the numerical values in Eq. (31) and with the numerical values in Eqs. (32) and (33) obtained for $\delta_2(s) = 0$, we summarize the following numerical values obtained for our final parameterization of $\delta(s)$:

$$a_0^0 = 0.220, \qquad \delta(M_K^2) = 40.4°, \qquad \delta(800) = 77.0°,$$

$$s_{Ad} = 0.41 M_\pi^2, \qquad r = (0.236 \text{ GeV}^2)\exp(i32.6°). \tag{50}$$

To analyze Eq. (49) further, we again parameterize $A(M_\pi^2)$ according to Eq. (37) to check the sensitivity of $g_{K\sigma}$ on $A(M_\pi^2)$ ; we get

$$g_{K\sigma}(y = 0) = \{-(72.0 \text{ MeV})\exp(-i74.7°)\}|A(M_K^2)| ,$$

$$g_{K\sigma}(y = +1/3) = \{-(74.5 \text{ MeV})\exp(-i78.6°)\}|A(M_K^2)| ,$$

$$g_{K\sigma}(y = -1/3) = \{-(67.9 \text{ MeV})\exp(-i66.2°)\}|A(M_K^2)| . \tag{51}$$

Comparison of Eq. (49) with Eq. (36) and of Eq. (51) with Eq. (38) shows that adding $\delta_2(s)$ to $\delta_1(s)$ has a rather small effect on $g_{K\sigma}$; also, just as we found for $g_{K\sigma 1}$, $g_{K\sigma}$ changes rather modestly around the central value $g_{K\sigma}(y = 0)$ as $A(M_\pi^2)$ is varied widely around the central value $|A(M_K^2)|/6$. Also, to get the results in Eqs. (49), (50) and (51), we chose $s_{90°} = (1100 \text{ MeV})^2$ as a compromise to try to match $|r|_{\text{ex}} = 0.218^{+0.023}_{-0.010}$ GeV$^2$ and $\delta(800)_{\text{ex}} = (82.3^{+10}_{-4})°$ of Eq. (31), and we got $|r| = 0.236$ GeV$^2$ and $\delta(800) = 77.0°$. If we change to $s_{90°} = (900 \text{ MeV})^2$ and repeat our calculations, we get
$|r| = 0.269$ GeV$^2$, $\delta(800) = 82.9°$ and $g_{K\sigma}(y = 0) = \{-(72.6 \text{ MeV})\exp(-i76.3°)\}|A(M_K^2)|$ compared with the result in Eq. (51). That is, the match to $|r|_{\text{ex}}$ deteriorates as the match to $\delta(800)_{\text{ex}}$ improves, but $g_{K\sigma}$ barely changes; this insensitivity to changes in $s_{90°}$ occurs primarily because an increase in $r$ and $g_{\sigma\pi\pi}$ is countered in Eq. (35) by a corresponding increase in $\Omega_2(s_\sigma, M_K^2)$ .

## 5. Discussion

Our simple parameterization of the $I=0$ $S$-wave $\pi\pi \to \pi\pi$ scattering phase shift $\delta(s) = \delta_1(s) + \delta_2(s)$ yields a complete simple solution for the relationship among $A(M_K^2)$, $A(M_\pi^2)$ and $g_{K\sigma}$. Our choice of numerical values for the parameters in this $\delta(s)$ gives an approximate numerical representation of the $\pi\pi \to \pi\pi$ amplitude for real $s$ ranging from the location of the sub-threshold or Adler zero at $s_{\text{Ad}} = 0.41 M_\pi^2$ to about $s = (800 \text{ MeV})^2$ and for $s$ near the sigma pole at $s = s_\sigma$. These numerical values of the parameters yield Eq. (49), repeated here,

$$g_{K\sigma} = -\{(84.0 \text{ MeV})\exp(-i88.7°)\} \times \{[(i1.60)\exp(i39.2°)]A(M_\pi^2) + |A(M_K^2)|\} , \tag{52}$$

as our final numerical result relating $A(M_K^2)$, $A(M_\pi^2)$ and $g_{K\sigma}$ .

The phase shift $\delta_1(s)$ alone with $\delta_2(s) = 0$ gives an adequate approximation to the $\pi\pi \to \pi\pi$ amplitude near $s = M_K^2$ and near $s = s_\sigma$ ; we introduced $\delta_2(s)$ to improve the approximation for $s$ near threshold and for higher $s$ near $(800 \text{ MeV})^2$. A comparison of $g_{K\sigma 1}$ of Eq. (36), obtained for $\delta_2(s) = 0$, with our final $g_{K\sigma}$ of Eq. (52) shows that introducing $\delta_2(s)$ results in only small changes to the numerical relationship. Moreover, the discussions following Eqs. (51) and (38) indicate that details of the high energy phase shift do not drastically affect the relationship among $A(M_K^2)$, $A(M_\pi^2)$ and $g_{K\sigma}$.

The results in Refs. [1] and [2] indicate that $A(M_\pi^2)$ is expected to be much smaller than $|A(M_K^2)|$, and we chose an approximate central value $A(M_\pi^2) = |A(M_K^2)|/6$ for which

$$g_{K\sigma}(y=0) = -\{(72.0 \text{ MeV})\exp(-i74.7°)\}|A(M_K^2)| \tag{53}$$

from Eq. (51). Although $A(M_\pi^2)$ may not be known precisely, Eq. (51) indicates a modest variation of $g_{K\sigma}$ around this central value for a very wide variation of $A(M_\pi^2)$. Therefore, a knowledge of $g_{K\sigma}$ is *practically* equivalent, via the central value of Eq. (53), to a knowledge of $A(M_K^2)$ and vice-versa.

If we assume the $\Delta I = 1/2$ rule and relate $A(M_K^2)$ to the amplitude $A(K_S^0 \to \pi^+\pi^-)$ for the decay $K_S^0 \to \pi^+ + \pi^-$, we get

$$g_{K\sigma}(y=0) = -\{(88.1 \text{ MeV})\exp(-i74.7°)\}|A(K_S^0 \to \pi^+\pi^-)| \tag{54}$$

from Eq. (53). This gives $|g_{K\sigma}(y=0)| = (88.1 \text{ MeV})|A(K_S^0 \to \pi^+\pi^-)|$ which is remarkably close to the result

$$|g_{K\sigma}| \simeq f_\pi |A(K_S^0 \to \pi^+\pi^-)|, \tag{55}$$

where $f_\pi = 92.2$ MeV is the pion decay constant, estimated in Ref. [13] on the basis of the linear sigma model.

Some time ago, Truong [14] used, and more recently [15] reviewed, a dispersion relation formalism in the variable $s_\text{Tr} = -k_K^2 = -(k_1 + k_2)^2$ without a spurion but, rather, with the kaon off the mass shell. We can rewrite his resulting $K \to 2\pi$ amplitude $A(s_\text{Tr})$ as

$$A(s_\text{Tr}) = \frac{s_\text{Tr} - M_\pi^2}{M_K^2 - M_\pi^2} A(M_K^2)\Omega(s_\text{Tr}, M_K^2), \tag{56}$$

which corresponds in form, *but not in principle*, to the second term in Eq. (4), that is, to $A(s)$ with $A(M_\pi^2)$ set to zero in Eq. (4). Therefore, if we simply set $A(M_\pi^2)$ to zero in Eq. (52), we get

$$g_{K\sigma,\text{Tr}} = -\{(84.0 \text{ MeV})\exp(-i88.7°)\}|A(M_K^2)| \tag{57}$$

for Truong's formalism, compared with our $g_{K\sigma}(y=0)$ of Eq. (53), $g_{K\sigma}(y)$ of Eq. (51) and $g_{K\sigma 1}(y)$ of Eq. (38). This suggests that a sufficiently reliable theoretical result for $g_{K\sigma}$ obtained from, say, a lattice calculation might distinguish between the spurion formalism that we used and the formalism with the kaon off the mass shell that Truong used.

We have not attempted a systematic evaluation of errors in our results, nor have we included inelasticity in $\pi\pi$ scattering. In view of our results, it may be desirable to undertake a more rigorous analysis and to include inelasticity, perhaps along the lines of Ref. [2].

## Acknowledgement

The author thanks H. Leutwyler for information on the residue $r$ of the sigma pole. .

## Appendix A. The slope $A'(M_\pi^2)$ of $A(s)$ at $s = M_\pi^2$

Suppose that, instead of Eq. (1), the dispersion relation for $A(s)$ is subtracted twice at $s = M_\pi^2$ such that

$$A(s) = A(M_\pi^2) + (s - M_\pi^2)A'(M_\pi^2)$$
$$+ \frac{(s - M_\pi^2)^2}{\pi} \int_{4M_\pi^2}^{\infty} \frac{\mathrm{Im}A(s')ds'}{(s' - M_\pi^2)^2(s' - s - i\varepsilon)} + (L.H.C.)$$

(A1)

where $A'(M_\pi^2)$ is the first derivative of $A(s)$ with respect to $s$ evaluated at $s = M_\pi^2$. Under the same assumptions as made in Sec. 2, the solution is

$$A(s) = \{A(M_\pi^2) + (s - M_\pi^2)[A'(M_\pi^2) - A(M_\pi^2)\Omega'(M_\pi^2, M_\pi^2)]\}\Omega(s, M_\pi^2) ,  \qquad (A2)$$

where $\Omega'(M_\pi^2, M_\pi^2)$ is the first derivative of $\Omega(s, M_\pi^2)$ with respect to s evaluated at $s = M_\pi^2$. Since a knowledge of $A'(M_\pi^2)$ would be desirable to calculate $A(s)$ from Eq. (A2), we now evaluate $A'(M_\pi^2)$ in terms of $g_{K\sigma}$ and $A(M_\pi^2)$. We can proceed by either

(i) evaluating the residue $\lim_{s \to s_\sigma}(s - s_\sigma)A(s)$ of Eq. (A2) for use with Eq. (8) and then isolating $A'(M_\pi^2)$, or
(ii) taking the derivative of Eq. (4) and then eliminating $|A(M_K^2)|$ by using $|A(M_K^2)|$ from Eq. (10). Here, we follow the second procedure; the first procedure gives the same result.

Taking the first derivative of Eq. (4) and evaluating the derivative at $s = M_\pi^2$ gives

$$A'(M_\pi^2) = \{[-1 + (M_K^2 - M_\pi^2)\Omega'(M_\pi^2, M_\pi^2)]A(M_\pi^2) + |A(M_K^2)|/|\Omega(M_K^2, M_\pi^2)|\}/(M_K^2 - M_\pi^2). \quad (A3)$$

Then, combining Eqs. (A3) and (10) to eliminate $|A(M_K^2)|$ gives

$$A'(M_\pi^2) = \frac{[-1 + (M_K^2 - M_\pi^2)\Omega'(M_\pi^2, M_\pi^2) + (s_\sigma - M_K^2)/(s_\sigma - M_\pi^2)]A(M_\pi^2)}{M_K^2 - M_\pi^2}$$
$$- \frac{g_{K\sigma}g_{\sigma\pi\pi}}{(s_\sigma - M_\pi^2)|\Omega(M_K^2, M_\pi^2)|\lim_{s \to s_\sigma}(s - s_\sigma)\Omega(s, M_\pi^2)\exp\{i\delta(M_K^2)\}} . \qquad (A4)$$

For the product form of $S(s)$ of Eq. (11), Eq. (A4) gives

$$A'(M_\pi^2) = \frac{[-1 + (M_K^2 - M_\pi^2)\{\Omega_1'(M_\pi^2, M_\pi^2) + \Omega_2'(M_\pi^2, M_\pi^2)\} + (s_\sigma - M_K^2)/(s_\sigma - M_\pi^2)]A(M_\pi^2)}{M_K^2 - M_\pi^2}$$

$$- \frac{g_{K\sigma}g_{\sigma\pi\pi}}{(s_\sigma - M_\pi^2)|\Omega_1(M_K^2, M_\pi^2)||\Omega_2(M_K^2, M_\pi^2)|[\Omega_2(s_\sigma, M_K^2)\exp\{i\delta_2(M_K^2)\}] \times R_{\Omega 1}} \,. \quad (A5)$$

Eq. (A4) gives $A'(M_\pi^2)$ for any $\delta(s)$, and Eq. (A5) gives $A'(M_\pi^2)$ for the product form of $S(s)$.

To evaluate $A'(M_\pi^2)$ numerically for the parameterizations in Secs. 3 and 4, the numerical values of the derivatives $\Omega_1'(M_\pi^2, M_\pi^2)$ and $\Omega_2'(M_\pi^2, M_\pi^2)$ are required. The derivatives of $\Omega_1(s, s_0)$ of Eq. (27) and $\Omega_2(s, s_0)$ of Eq. (44) evaluated at $s = M_\pi^2$ give

$$\Omega_1'(M_\pi^2, M_\pi^2) = \frac{1}{4M_\pi^2}\left\{\frac{1}{-3/4 + \gamma^2} + \frac{\alpha[1 - (9/4)\beta]}{\sqrt{3} + (3/2)\alpha[1 - (3/4)\beta]}\right\},$$

$$\Omega_2'(M_\pi^2, M_\pi^2) = \frac{1}{4M_\pi^2}\left\{\frac{1}{-3/4 + \omega^2} + \frac{-\sqrt{3}\xi + \lambda[1 - (9/4)\zeta]}{\sqrt{3}[1 - (3/4)\xi] + (3/2)\lambda[1 - (3/4)\zeta]}\right\}. \quad (A6)$$

The numbers for the parameters used in Secs. 3 and 4 give

$$\Omega_1'(M_\pi^2, M_\pi^2) = 0.619/(M_K^2 - M_\pi^2) \,, \qquad \Omega_2'(M_\pi^2, M_\pi^2) = -0.000604/(M_K^2 - M_\pi^2) \,. \quad (A7)$$

Then, Eq. (A5) with other numerical values taken from Secs. 3 and 4 gives

$$A'(M_\pi^2)_1 = \{(3.75 \text{ GeV}^{-2})\exp(-i70.9°)\}A(M_\pi^2) - \{(37.0 \text{ GeV}^{-3})\exp(i85.5°)\}g_{K\sigma 1} \,, \quad (A8)$$

$$A'(M_\pi^2) = \{(3.75 \text{ GeV}^{-2})\exp(-i71.0°)\}A(M_\pi^2) - \{(34.1 \text{ GeV}^{-3})\exp(i88.7°)\}g_{K\sigma} \,. \quad (A9)$$

The subscript 1 on $A'(M_\pi^2)_1$ in Eq. (A8) indicates the approximation $\delta_2(s) = 0$; Eq. (A9) gives our final result for $\delta(s) = \delta_1(s) + \delta_2(s)$.